\documentclass[sigconf, authorversion]{acmart}

\usepackage{fvextra}
\usepackage{fancyvrb}
\usepackage{subfig}
\usepackage{listings}
\usepackage{url}

\AtBeginDocument{%
  \providecommand\BibTeX{{%
    \normalfont B\kern-0.5em{\scshape i\kern-0.25em b}\kern-0.8em\TeX}}}

\setcopyright{acmcopyright}
\copyrightyear{2023}
\acmYear{2023}
\acmDOI{10.1145/3624062.3624215}

\setcopyright{acmlicensed}\acmConference[SC-W 2023]{Workshops of The International Conference on High Performance Computing, Network, Storage, and Analysis}{November 12--17, 2023}{Denver, CO, USA}
\acmBooktitle{Workshops of The International Conference on High Performance Computing, Network, Storage, and Analysis (SC-W 2023), November 12--17, 2023, Denver, CO, USA}

\acmConference[SC '23]{The International Conference for High Performance Computing, Networking, Storage, and Analysis}{November 12--17, 2023}{Denver, CO}
\acmPrice{15.00}
\acmISBN{979-8-4007-0785-8/23/11}
\acmSubmissionID{153}

\begin{document}
\title{BGLS: A Python Package for the Gate-by-Gate Sampling Algorithm to Simulate Quantum Circuits}

\author{Alex Shapiro}
\email{alexander.shapiro@epfl.ch}
\orcid{0009-0007-6754-4845}
\affiliation{%
  \institution{EPFL}
  \city{Lausanne}
  \country{Switzerland}
}

\author{Ryan LaRose}
\email{rmlarose@msu.edu}
\orcid{0000-0002-5398-3341}
\affiliation{%
  \institution{Michigan State University}
  \city{East Lansing}
  \country{USA}
}

\begin{abstract}
    The classical simulation of quantum computers is in general a computationally {hard} problem. To emulate the behavior of realistic devices, it is sufficient to sample bitstrings from circuits. Recently, Ref.~\cite{Bravyi_2022} introduced the so-called gate-by-gate sampling algorithm to sample bitstrings and showed it to be computationally favorable in many cases. Here we present \Verb|bgls|, a Python package which implements this sampling algorithm. \Verb|bgls| has native support for several states and is highly flexible for use with additional states. We show how to install and use \Verb|bgls|, discuss optimizations in the algorithm, and demonstrate its utility on several problems.
\end{abstract}



\received{September 18 2023}

\maketitle

\section{Introduction}
The simulation of quantum circuits with classical computers is a fundamental tool in the rapidly growing field of quantum computing. It is important for prediction and verification of results obtained on physical hardware, as well as for enabling agile prototyping of quantum algorithms and the circuits implementing them. It continues to be increasingly useful as circuit sizes under study grow.

The state of a circuit can be fully described at any time by its wavefunction $\Psi$, but there is no efficient way to directly recover this information. Instead, we must content ourselves with measurement samples. For an $n$-qubit system, such measurements consist of length $n$ bitstrings i.e. $b_0b_1\dots b_n$ with each $b_i \in \{0,1\}$, corresponding to projective measurement of $\Psi$ onto the computational basis. A given bitstring is thus sampled with frequency following its corresponding amplitude component of our state in this basis as per $|\langle b_0\dots b_n | \Psi\rangle|^2$. Hence, classical simulation only needs to provide this so-called \textit{weak} simulation to capture realistic behavior. Importantly, it is such sampling (on random circuits) that is used as a classification of ``quantum supremacy''\cite{Bouland2019}.

Unfortunately, simulating this sampling is inherently limited and is widely known to be \#P hard~\cite{Bravyi_2022}. Improvements to this sampling task are thus essential to expanding the reach of what is classically simulable. Excitingly, an alternative to the standard algorithm for simulating sampling from quantum circuits has been proposed by Bravyi, Gosset, and Liu~\cite{Bravyi_2022} which can offer significantly improved performance for several problems.

Here we introduce \Verb|bgls| (GitHub: ~\cite{Shapiro_BGLS_2023}), a Python package providing this sampling routine. It directly interfaces with Google's Cirq~\cite{cirq_developers_2022_7465577} framework and is built to be highly agnostic - functioning on essentially any arbitrary quantum state representation. To this end, it additionally provides many tools for working with states benefiting from this algorithm, most notably (near-)Clifford stabilizer states and matrix product states. 

The remainder of the paper is structured as follows. In Sec.~\ref{sec:gate-by-gate-sampling-algorithm} we begin with an explanation of the sampling algorithm underlying \verb|bgls|, a comparison with the traditional method, and point towards cases where \verb|bgls| is favorable. In Sec.~\ref{sec:core-features-of-bgls} we explain the structure of the \verb|bgls.Simulator| class and show how to use it with a basic example, as well as highlighting other core features. In Sec.~\ref{sec:examples-using-bgls} we demonstrate several examples using \verb|bgls| including (near-)Clifford states, matrix product states, and an implementation of the Quantum Approximate Optimization Algorithm~\cite{farhi2014quantum}. 

\section{Background: The Gate-by-Gate Sampling Algorithm} \label{sec:gate-by-gate-sampling-algorithm}

Before introducing \verb|bgls|, we first summarize the gate-by-gate sampling algorithm from~\cite{Bravyi_2022}.

To start with general notation, we consider the task of generating samples from a quantum circuit consisting of $n$ qubits and $d$ gates. The system is assumed to start in the computational basis 0 state, i.e. $|\Psi_0\rangle = |0...0\rangle$, and our circuit consists of a sequence of gates $U=U_dU_{d-1}\dots U_1$. The final state of the system is given simply by $|\Psi_f\rangle = U_dU_{d-1}\dots U_1|\Psi_0 \rangle$. The algorithm's output should be a bitstring drawn from the probability distribution $P(b_0\dots b_n) = |\langle b_0\dots b_n | \Psi_f \rangle|^2$. 

Conventionally, the final state $\Psi_f$ is first computed and then marginal distributions on each qubit are computed in order to sample a bitstring. This \textit{qubit-by-qubit} sampling algorithm can be sketched as follows: \begin{enumerate}
    \item Initialize and fully run the circuit to obtain $|\Psi_f \rangle$.
    \item Sequentially measure each qubit by computing and sampling from its \textit{marginal probability distribution} given the preceding measurement values. 
    \item After sampling all qubits, the final bitstring measurement is returned.
\end{enumerate}
This procedure thus requires one to compute marginal probabilities of the system $n$ times. 

We now compare this with the \textit{gate-by-gate} sampling algorithm. The idea is that rather than first fully evolving the circuit and then sequentially measuring qubits, it instead walks through the circuit one gate at a time, sampling from the intermediate output distribution $P(b0\dots b_n)$ at each step. It is this substitution of computing bitstring probabilities rather than marginal distrubtions that is the key difference. The algorithm can be sketched as follows:
\begin{enumerate}
    \item Initialize a bitstring $b = 0\dots 0$.
    \item Loop over gates in the circuit. For each:
    \begin{itemize}
        \item The \textit{support} is the set of qubits acted on by the current gate. Consider all possible candidate bitstrings generated from the current $b$ that vary over the support while fixing the remaining indices.
        \item Apply the current gate to update the state $|\Psi \rangle$.
        \item For each candidate bitstring, compute its probability of being measured given the current state. Sample from this distribution to update $b$.
    \end{itemize}
    \item After the final gate, the resultant bitstring is returned as the measurement result.
\end{enumerate}
As in~\cite{Bravyi_2022}, let the cost of computing a bitstring probability from an $n$-qubit, depth $d$ circuit be $f(n,d)$. One expects the cost of computing marginal distributions to be comparable to $f(n, 2d)$, so the gate-by-gate sampling algorithm provides an enhancement on the order of ${f(n,2d)} / {f(n,d)}$ compared to the qubit-by-qubit sampling algorithm.

\section{The BGLS Package} \label{sec:core-features-of-bgls}

BGLS is available on PyPI and can be installed via \begin{Verbatim}
    pip install bgls
\end{Verbatim} 
The code is hosted on GitHub at \href{https://github.com/asciineuron/bgls}{github.com/asciineuron/bgls} and the documentation is available at \href{https://asciineuron.github.io/bgls/}{asciineuron.github.io/bgls/}.

\subsection{Core Usage}
The core object of BGLS is the \verb|bgls.Simulator| which inherits from the \verb|cirq.Simulator| and so provides an identical interface to simulating circuits in Cirq. To construct a \verb|bgls.Simulator| one needs (1) an \verb|initial_state| of any generic \verb|State| type representing the quantum state, (2) a function \verb|apply_op| to apply a \verb|cirq.Operation| to a \verb|State| and thus update it, and (3) a \verb|compute_probability| function yielding the probability of measuring a particular bitstring from a given \verb|State|.


Basic usage, taking advantage of underlying Cirq structures for working with a state vector representation, can be seen in the following snippet.
\begin{Verbatim}[fontsize=\small]
import cirq
import bgls

nqubits = 2
qubits = cirq.LineQubit.range(nqubits)
circuit = cirq.Circuit(
    cirq.H.on(qubits[0]),
    cirq.CNOT.on(qubits[0], qubits[1]),
    cirq.measure(*qubits, key="z")
)

simulator = bgls.Simulator(
    initial_state=cirq.StateVectorSimulationState(
        qubits=qubits, 
        initial_state=0),
    apply_op=cirq.protocols.act_on,
    compute_probability=bgls.born.compute_probability\
                                  _state_vector,
)
results = simulator.run(circuit, repetitions=10)
cirq.plot_state_histogram(results)
\end{Verbatim}

\begin{figure}
    \centering
    \includegraphics[width=0.5\linewidth]{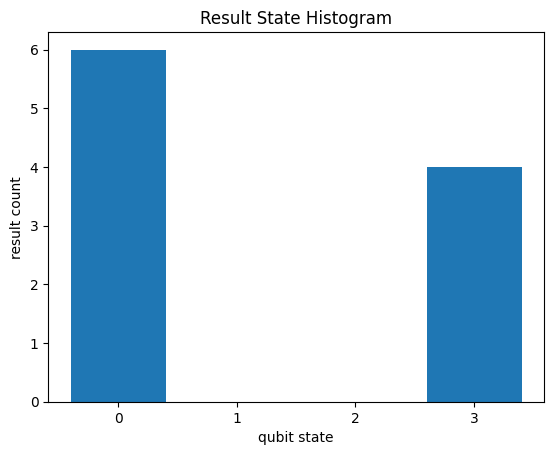}
    \caption{Measurement results for a simple ghz circuit.}
    \label{fig:statevec-ghz}
\end{figure}

The output of this code block is shown in Fig.~\ref{fig:statevec-ghz}. We emphasize again that the \verb|bgls.Simulator| provides an identical interface to simulating circuits in Cirq, in particular through measurement keys, output format, parametric support (see Sec.~\ref{sec:qaoa} for an example), etc.

\subsection{Features of BGLS}

\subsubsection{Support for non-unitary operations including noise and mid-circuit measurements}

The gate-by-gate sampling algorithm can be easily modified if some operations in $U=U_d\dots U_1$ are non-unitary. BGLS supports this case as long as the \verb|apply_op| function provided to the \verb|bgls.Simulator| can decompose operations - this task is particularly easy in Cirq and an example is discussed in the documentation at \href{https://asciineuron.github.io/bgls/features.html}{https://asciineuron.github.io/bgls/features.html}. In this case the method of quantum trajectories is applied to simulate non-unitary operations. Two common use cases of non-unitary operations are noisy simulation and mid-circuit measurement and these are fully supported in BGLS. The only difference relative to unitary circuits is that automatic sample parallelization (Sec.~\ref{sec:sample-parallelization}) typically cannot be applied in quantum trajectories since the number of wavefunctions is much larger than the number of samples.

\subsubsection{Circuit optimization}

BGLS provides support to optimize circuits for the gate-by-gate sampling algorithm through the function \verb|bgls.optimize_for_bgls|. As a simple but illustrative example, consider a circuit with five sequential single-qubit operations. The circuit optimization will merge these operations into one single-qubit operation so that the bitstring only gets updated once instead of five times. An example is provided in the BGLS documentation at \href{https://asciineuron.github.io/bgls/tips.html}{asciineuron.github.io/bgls/tips.html}. As shown there, circuit optimization on random eight qubit circuits with up to 50 layers of operations leads to a runtime improvement between 1.5x and 2x.

\subsubsection{Automatic sample parallelization} \label{sec:sample-parallelization}

\begin{figure}
    \centering
    \includegraphics[width=0.75\linewidth]{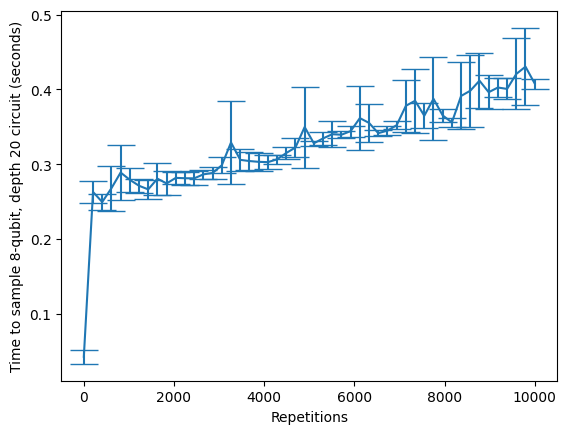}
    \caption{Sample parallelization saturates runtime at many repetitions.}
    \label{fig:measurement-parallelization}
\end{figure}

While the gate-by-gate sampling algorithm as presented in Sec.~\ref{sec:gate-by-gate-sampling-algorithm} updates one bitstring per evolution of $\Psi$, all bitstrings can be updated in parallel when $\Psi$ is the same for all bitstrings.
The basic idea is that for generating \textit{N} samples, a list of \textit{N} bitstrings is created, and at every step all are independently updated as expected. To further improve performance, \verb|bgls| represents this as a dictionary mapping each current unique bitstring \textit{b} to its multiplicity \textit{m}. At each gate we then determine all candidates for every \textit{b} and draw \textit{m} updated samples. Sampling performance can be seen in Fig.~\ref{fig:measurement-parallelization}. In particular, runtime saturates at large repetitions as there are at most $2^n$ unique bitstrings for an $n$ qubit system, limiting the total dictionary size.

For circuits with non-unitary operations, the final state is stochastic and must be rerun for each sample desired. Additionally as shall be described below, sampling near-Clifford circuits using \verb|act_on_near_clifford| must be repeated to fully explore the space.

\subsubsection{Usage with non-Cirq circuits}

BGLS only directly supports Cirq circuits, but most circuit classes from quantum software packages can convert to and from Cirq circuits. For example one can convert a QASM circuit~\cite{cross2017open} via \verb|from cirq.contrib.qasm_import|~\cite{cirq_developers_2022_7465577_interop}. An example is shown at \href{https://asciineuron.github.io/bgls/features.html}{asciineuron.github.io/bgls/features.html}

\section{Examples using BGLS} \label{sec:examples-using-bgls}
\subsection{Clifford Circuits and Stabilizer States}
\subsubsection{Background and theory of Clifford circuits}
We first briefly review some core concepts necessary to understanding Clifford circuits and stabilizer states. This is by no means comprehensive and many sources can be found elsewhere.

First is the definition of the \textit{Pauli group}. For a system of $n$-qubits, the Pauli group $G_n$ is a matrix group consisting of all possible applications of I and the Pauli matrices X, Y, and Z, along with prefactors $\pm 1$, $\pm i$ to each of the qubits' Hilbert spaces in the tensor product. As an example, for 1 qubit this set consists of $\{\pm I, \pm i I, \pm X, \pm i X, \pm Y, \pm i Y, \pm Z, \pm i Z\}$.

Given this, an $n$-qubit stabilizer state $|\Psi \rangle$ is any state such that there exists a subgroup $S \le G_n$ of size $2^n$ that \textit{stabilizes} the state, that is for every $s \in S$, $s|\Psi \rangle = |\Psi \rangle$. For example, if $|\Psi\rangle = 1/\sqrt 2 (|0\rangle + |1\rangle)$, we see it is stabilized exactly by I and X. In particular, an additional consequence is that stabilizer states are exactly those which are reachable from $|0^n\rangle$ by applying exclusively the so-called \textit{Clifford} gates. These consist of I, H, S, CNOT, and any combination thereof (e.g. they thus include the Pauli gates from combinations of S and H).

A final important reminder is that these states fall under the \textit{Gottesmann-Knill theorem} \cite{gottesman1998heisenberg}. This states that any circuit consisting only of initial basis state preparation, Clifford gate applications, and measurement of the qubits in the computational basis is able to be efficiently simulated classically.

\subsubsection{Representation by CH-Form}
To be able to work with such circuits, the first ingredient our \verb|bgls| sampler needs is a concrete representation of the state. For this, we turn to the so-called "CH-form" representation \cite{Bravyi_2019}, which is itself an extension of the ``stabilizer tableaux'' \cite{Aaronson_2004}. An existing implementation of this can be found via \verb|cirq.sim.StabilizerChFormSimulationState|, which \verb|bgls| \\builds upon to provide necessary tooling. We give a brief recap of the structure of this representation and how bitstring amplitudes can be computed.

The original paper showed that in general, any stabilizer state can be expressed as $|\Psi\rangle = \omega U_CU_H|s\rangle$, where $U_c$ is a tensor product of some number of S and CNOT gates, and $U_H$ is a product of H and I gates. $\omega$ is a complex scalar, and $|s\rangle \in \{0,1\}^n$ \cite{Bravyi_2019}.

The state of the system was specified with $n\times n$ binary matrices F, G, M, a phase $\gamma\in \mathbf{Z}_4^n$, and $v\in \{0,1\}^n$. The equations relating these to $U_C, U_H$ are: 
\begin{displaymath}
    U_C^{-1}Z_pU_C = \prod_{j=1}^nZ_j^{G_{p,j}}
\end{displaymath}
\begin{displaymath}
    U_C^{-1}X_pU_C=i^{\gamma_p}\prod_{j=1}^nX_j^{F_{p,j}}Z_j^{M_{p,j}}
\end{displaymath}
\begin{displaymath}
    U_H = H_1^{v_1}\otimes H_2^{v_2} \otimes\dots\otimes H_n^{v_n}
\end{displaymath}

Cirq's implementation directly stores these variables, and wraps handling of gate application with the native \verb|act_on| protocol. Lastly, we need to be able to compute the probability of an arbitrary bitstring $P(b)$ for such a state. The corresponding bitstring \textit{amplitude} was given, up to some $\mu\in\mathbf{Z_4}$ as:
\begin{displaymath}
    \langle b | \Psi \rangle = 2^{-|v|/2}i^\mu \prod_{J:v_j=1}(-1)^{(bF)_js_j}\prod_{j:v_j=0}\langle (bF)_j|s_j \rangle
\end{displaymath}
This is found in \verb|StabilizerChFormSimulationState|'s \\ \verb|inner_product_of_state_and_x| function, from which computation of the probability is trivial. Of note is the computational complexity for this class of states. The cost of computing a desired probability was originally derived \cite{Bravyi_2019} to be $O(n^2)$ for $n$ qubits, which is \textit{polynomial} rather than exponential, and notably independent of the circuit depth. From this we can see that $f(n,d)=O(dn^2)$.

\subsubsection{Pure Clifford results}
Given the above framework, we turn to our implementation and support of this in \verb|bgls|. Specifically, the package provides the \verb|compute_probability_stabilizer_state| function. This operates on any valid \\\verb|StabilizerChFormSimulationState| and calculates via the above the measurement probability of a desired bitstring.

\begin{figure}%
    \centering
    \subfloat[\centering Runtime scaling with depth.]{{\includegraphics[width=0.45\linewidth]{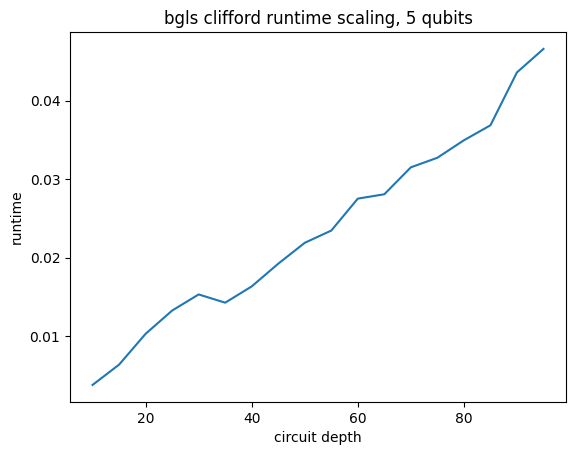} }}%
    \qquad
    \subfloat[\centering Runtime scaling with width.]{{\includegraphics[width=0.45\linewidth]{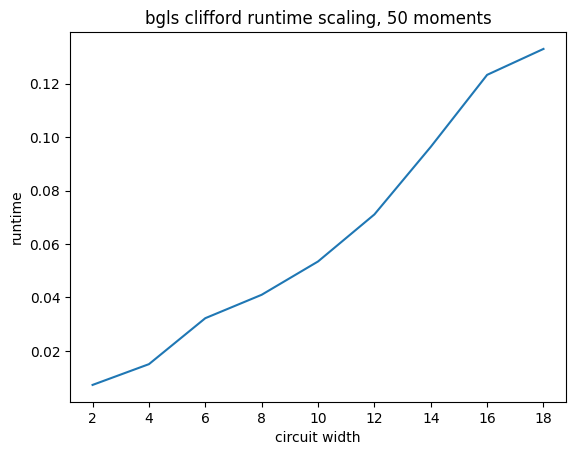} }}%
    \caption{Comparison of sampling runtime for Clifford circuits as depth or width is varied.}
    \label{fig:clifford-scaling}
\end{figure}

We are able to investigate the runtime scaling of sampling from such a state by constructing random circuits with the \verb|bgls| \\\verb|generate_random_circuit| function. This is derived from the Cirq equivalent, but provides simpler specification of the target gate set. We construct random circuits consisting exclusively of H, S, and CNOT gates, and either vary the depth of the circuit or the number of qubits in the system. Observed in Fig.~\ref{fig:clifford-scaling} is characteristic runtime scaling, confirming expected behavior.

It can be seen that the gate-by-gate sampling algorithm for these states is of an equivalent computational complexity as traditional sampling, and offers no direct benefit. However, the framework for handling these stabilizer states allows sampling from much more promising \textit{near-Clifford} circuits as we discuss in the following section.

\subsection{Near-Clifford and the Sum-Over-Clifford Technique}
\subsubsection{Background and review}
Importantly, pure Clifford circuits and stabilizer states are inherently limited as the Clifford gate set is not universal. That is, it is not necessarily possible to approximate an arbitrary circuit solely in terms of these gates. Commonly used to remedy this is the Toffoli gate, which together with the Clifford group forms a universal gate set \cite{kliuchnikov2013synthesis}. This no longer satisfies the Gottesman-Knill theorem, hence such circuits are not necessarily efficiently simulable. However, in many cases only relatively few T gates are needed to express a circuit. This allows leveraging the stabilizer state framework on a much larger class of circuits, and forms the basis of the \textit{sum-over-Cliffords} technique presented in the same paper as the CH-representation \cite{Bravyi_2019}. We first briefly recap their results, and then show its implementation in \verb|bgls|.

Their idea is that any non-Clifford gate $U$ can be decomposed into a sum of Clifford gates $K$ as $U=\sum_i c_i K_i$. This was quantified by the ``stabilizer extent'' $\zeta$ which is the minimum norm of all such possible decompositions, and gives a heuristic of how non-Clifford the system is.

In particular it was shown that any diagonal rotation gate $R(\theta)=e^{-iZ\theta/2}$ is ideally decomposed as
\begin{displaymath}
    R(\theta)=(\cos(\theta/2)-\sin (\theta/2))I + \sqrt{2}e^{-i\pi/4}\sin (\theta/2)S
\end{displaymath}
where $S$ is the Phase gate $S=T^2$. Notably, T can be equivalently written as $R(\pi/4)$ and hence expanded in this way.

\subsubsection{Implementation in BGLS}
Taking advantage of this, we implement it in \verb|bgls| with the novel function \verb|act_on_near_clifford| for gate application, which enables sampling from general Clifford+$R_Z(\theta)$ circuits by simulating with a stabilizer state. This functionality is new to Cirq, where the stabilizer state simulator is limited to pure Clifford circuits.

Its method of operation is straightforward: for any gate with a stabilizer effect checked via \verb|cirq.has_stabilizer_effect|, the standard \verb|act_on| is applied. For a $R_Z(\theta)$ gate (a subclass of \\ \verb|cirq.ops.common_gates.ZPowGate|), we extract $\theta$, compute the relative probabilities of I and S by the above equation, and choose one following this distribution to substitute for $R$.

For a circuit with $N$ such $R(\theta)$ gates, there are clearly $2^{N}$ terms in the final expansion. Hence, for a given sample we are stochastically exploring only one of these branches. In the following section we investigate the results of using stabilizer states on such near-Clifford circuits.

\subsubsection{Near-Clifford results}
\begin{figure}%
    \centering
    \subfloat[\centering Overlap with increasing runtime for pure-Clifford and near-Clifford.]{{\includegraphics[width=0.45\linewidth]{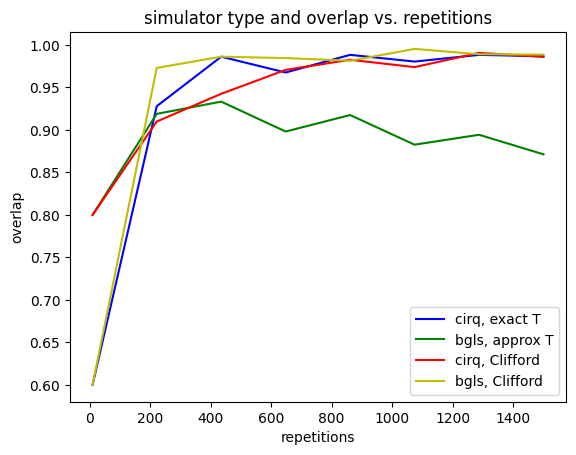} }}%
    \qquad
    \subfloat[\centering Clifford+$R(\theta)$ overlap]{{\includegraphics[width=0.45\linewidth]{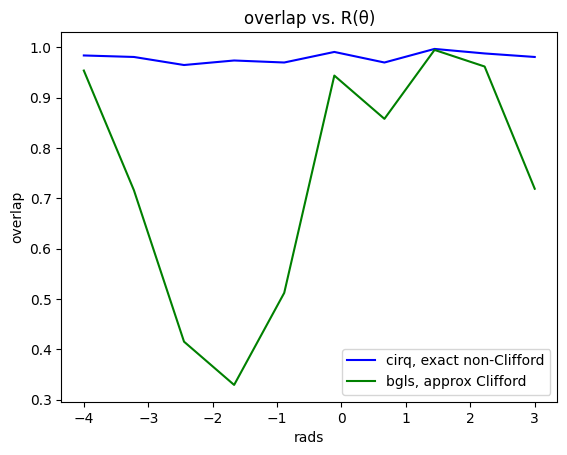} }}%
    \caption{Overlap attained with sum-over-Cliffords sampling.}
    \label{fig:clifford-sum-overlap}
\end{figure}

What follows can similarly be found with implementation details on the project's homepage. As a first analysis, we compare the accuracy of simulating near-Clifford and pure-Clifford circuits. A random circuit is constructed consisting of Clifford+T gates, and a copy is made where each T gate is substituted by S. We sample from these using both the \verb|bgls| stabilizer state simulator, as well as an exact \verb|cirq.Simulator|, and plot in Fig.~\ref{fig:clifford-sum-overlap} the fractional overlap with what we would expect from the ideal distribution. Observed is a noticeable lag for \verb|bgls| applied to non-Clifford circuits, i.e. using the sum-over-Cliffords technique. This arises from the $2^{\# T}$ stabilizer states required to represent the non-Clifford circuit, and hence a given number of samples will explore a correspondingly smaller portion of the output distribution.

Next, we explore the effect rotation angle has on output fidelity. We take a fixed random Clifford+T circuit and varying $\theta$, substitute all T with $R(\theta)$. Again the overlap with the ideal distribution is plotted for a fixed number of samples, both for an exact simulator and our sum-over-Cliffords implementation. We see in Fig.~\ref{fig:clifford-sum-overlap} an overlap for \verb|bgls| that greatly fluctuates with angle used. This suggests that perhaps $R(\theta)$ near maxima of this plot could provide a more efficient alternative to T gates in achieving a universal gate set, and could be an avenue for future investigation.

\begin{figure}
    \centering
    \includegraphics[width=0.5\linewidth]{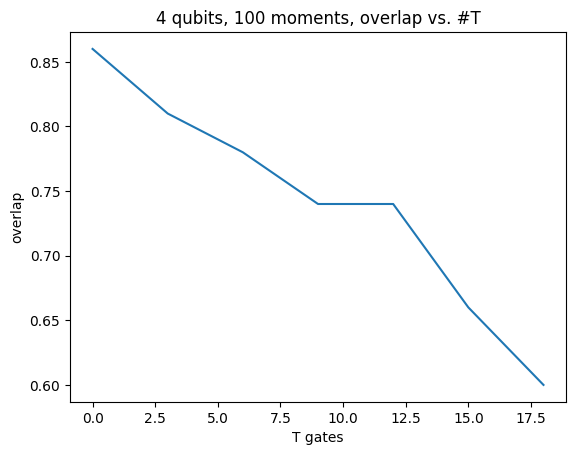}
    \caption{Sum-over-Clifford overlap decreases with additional T gates.}
    \label{fig:clifford-sum-add-T}
\end{figure}

As a last comment about such near-Clifford simulation techniques, we alternatively examine the effect of T gates by instead constructing a random pure-Clifford circuit of 100 moments and progressively replace more gates with T. For each circuit we plot overlap for a fixed number of samples. As the circuit becomes increasingly non-Clifford, we observe in Fig.~\ref{fig:clifford-sum-add-T} a decrease in overlap. This highlights the namesake of \textit{near}-Clifford, in that adequate performance is limited by the degree in which the circuit is non-Clifford.

\subsection{Matrix Product States}
\subsubsection{Introduction}
Perhaps the most promising state representations to use with \verb|bgls| are those based on matrix product states (MPS), and we provide full support and examples for working with these. Such states have been of great interest to researchers in recent years, and there is much relevant literature for readers interested in more comprehensive information \cite{perezgarcia2007matrix, biamonte2017tensor}. Here, we focus our attention on the specific MPS form presented in the paper ``What limits the simulation of quantum computers?'' \cite{Zhou_2020}.

In their notation, any $n$-qubit state can be written in the tensor network form:
\begin{displaymath}
    |\Psi\rangle = \sum_{i_1\dots i_N}\sum_{\mu_1\dots \mu_N} M(1)^{i_1}_{\mu_1}\dots M(n)^{i_n}_{\mu_n}|i_1\dots i_n\rangle
\end{displaymath}
Each $i_j \in \{0,1\}$ and hence we can see that this is a sum of tensor product terms over possible computational basis (i.e. bitstring) states. There is a unique tensor $M(j)$ corresponding to each qubit. For each of these, the tensor dimensions are given by $i_j$ of size 2, corresponding to the given qubit's 0 and 1 components, and $\mu_j\in\{1,\dots, \chi_j\}$ of size $\chi_j$. $\chi$ is referred to as the local dimension, and is a quantity that increases with and controls the degree of entanglement with other qubits.

It is thus evident that retrieving any quantity of interest (here again bitstring amplitudes) necessitates expensive contraction of this tensor network. As we saw, $\chi$ dominates the size of these tensors, and thus matrix product states are most useful where the total degree of entanglement is limited.

\subsubsection{Implementation in BGLS}
Handily, this structure is already implemented to interface with Cirq via the experimental \\\verb|cirq.contrib.quimb.MPSState| class. We next examine the data structures and show how specific bitstring amplitudes can be computed, a feature not present in the existing implementation.

In particular, a \verb|MPSState| stores a list of $n$ quimb \cite{Gray2018} tensors \verb|M| representing the above. Upon investigation, we see each is initially of shape $(2,)$ (this corresponds to the $i_j$ dimension), and for each additional qubit it is entangled with by multi-qubit gates, it gains another axis of dimension 2. An \verb|MPSState| is able to compute the full state vector by completely contracting this tensor network.

\verb|bgls| improves upon this by allowing calculation of \textit{specific} bitstring amplitudes, a feature not natively provided by the \verb|MPSState| class. By considering only a single bitstring, at each qubit we can eliminate the first tensor axis, choosing the index matching the desired bitstring's value. This tensor subset is extracted with the quimb \verb|isel| function, where the appropriate tensor label is provided with the \verb|MPSState|'s \verb|i_str| function. Proceeding this way through all $n$ tensors, a subnetwork is created of much smaller size. The desired amplitude is then the result of full contraction of this \textit{reduced} network, and is much less computationally intensive. This algorithm and the necessary \verb|quimb| functions can be understood as follows: 
\begin{Verbatim}[samepage=true]
def mps_bitstring_probability(mps, btstr):
    M_sub = []
    for i, Ai in enumerate(mps.M):
        qindx = mps.i_str(i)
        A_sub = Ai.isel({qindx: int(btstr[i])})
        M_sub.append(A_sub)
    tn = qtn.TensorNetwork(M_subset)
    st = tn.contract(inplace=False)
    return np.power(np.abs(st), 2)
\end{Verbatim}

A more complete walkthrough with graphical representation can be found in \verb|bgls|'s documentation \cite{Shapiro_BGLS_2023}.

\subsubsection{MPS results}
Armed with a state representation and bitstring probability function, we are now ready to use \verb|bgls| with MPS for circuit sampling. As a naive example, we consider a GHZ circuit with randomly sequenced CNOTs \cite{Gray2018}. Importantly, GHZ states can be \textit{analytically} represented trivially with MPS due to their symmetries. However, when blindly simulating these circuits this example becomes particularly hard as GHZ states are \textit{maximally} entangled. 

\begin{figure}%
    \centering
    \subfloat[\centering Random GHZ circuit]{{\includegraphics[width=0.45\linewidth]{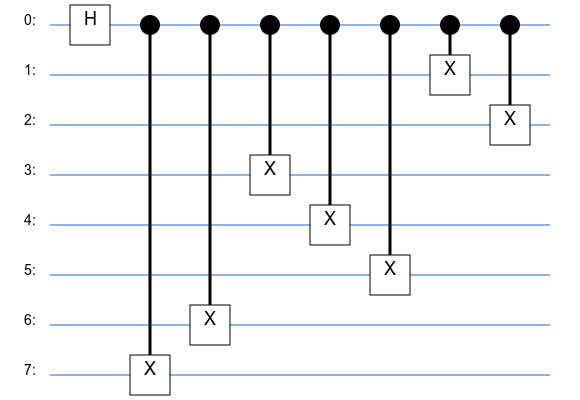} }}%
    \qquad
    \subfloat[\centering Runtime scaling]{{\includegraphics[width=0.45\linewidth]{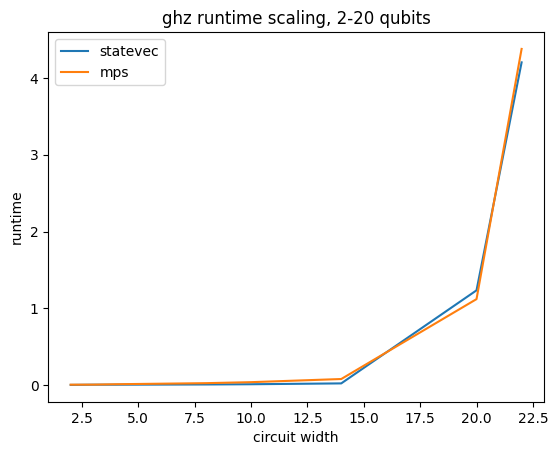} }}%
    \caption{Runtime scaling of mps compared to a state vector for a randomly-connected GHZ circuit of increasing width.}
    \label{fig:ghz-runtime-mps}
\end{figure}

Quantifying this, we construct such random GHZ circuits of width 2-20 and perform sampling with both the MPS and state vector representations. A characteristic circuit and a plot of the runtimes is seen in Fig.~\ref{fig:ghz-runtime-mps}. Observed is exponential runtime scaling of \textit{both} representations. This demonstrates MPS's exponential scaling with entanglement, and emphasizes that one needs particular care to achieve full performance from tensor network states. 

\begin{figure}%
    \centering
    \subfloat[\centering Random circuit mps vs state vector runtime scaling.]{{\includegraphics[width=0.45\linewidth]{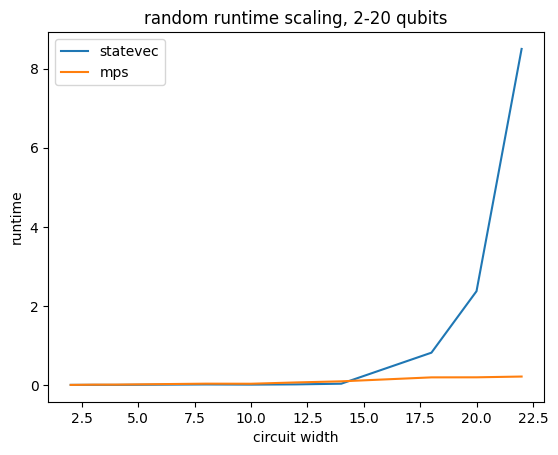} }}%
    \qquad
    \subfloat[\centering Runtime scaling for fixed CNOTs.]{{\includegraphics[width=0.45\linewidth]{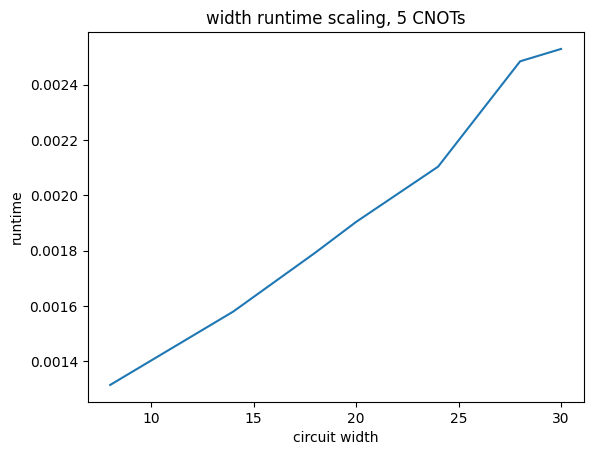} }}%
    \caption{For random circuits of increasing width, mps sampling is much more efficient than with a state vector. For a fixed degree of entanglement, scaling is linear in circuit width.}
    \label{fig:mps-random-runtime}
\end{figure}

Repeating the same procedure but instead with a \textit{random} circuit of fixed depth at each width, we see in Fig.~\ref{fig:mps-random-runtime} that using a MPS representation has drastically reduced runtime compared to a state vector. In this case, only a subset of gates are multi-qubit and entanglement producing. Hence for a shallow enough circuit the degree of entanglement lags behind the maximum (which exponentially increases with width). We could expect that for deep enough random circuits of a given width, all qubits are eventually entangled and MPS performance gains would give way.

Lastly, we consider random circuits of increasing width consisting of 1-qubit gates and a fixed number of 2-qubit CNOT gates. We observe in Fig.~\ref{fig:mps-random-runtime} a near linear sampling runtime scaling. Taken together, these results corroborate the $O(n\chi ^3)$ computational complexity of calculating amplitudes as was originally estimated \cite{Zhou_2020}.

\subsection{Example: MPS applied to QAOA} \label{sec:qaoa}
Taking this infrastructure, we next turn to a practical example demonstrating \verb|bgls|'s usefulness. In particular, a common task to perform on quantum computers is the ``Quantum Approximate Optimization Algorithm'' \cite{farhi2014quantum}. The idea behind this technique is to take a computationally hard optimization problem and map it onto Hamiltonians such that it can be represented by a circuit via unitaries $U=e^{i H}$. In this case qubit bitstrings map onto states of the original problem, and after optimizing the parameters of the QAOA circuit to maximize/minimize the expectation value of the corresponding cost function, we can expect resultant bitstring measurements to correspond to solutions of the classical problem.

Here we consider QAOA applied to the MaxCut problem. This is an NP-hard problem on graphs, where the objective is to find a partition of the vertices into two sets such that the number of edges being cut (i.e. its vertices belong to either set) is maximized. 

\begin{figure}%
    \centering
    \subfloat[\centering A characteristic sparse random graph.]{{\includegraphics[width=0.45\linewidth]{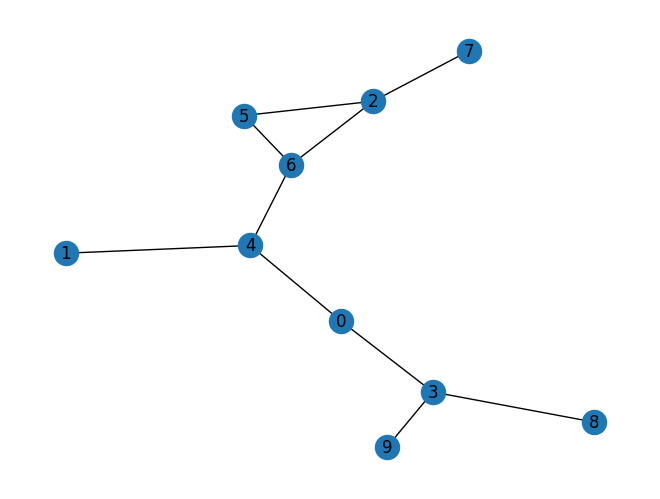} }}%
    \qquad
    \subfloat[\centering The corresponding QAOA circuit for MaxCut.]{{\includegraphics[width=0.45\linewidth]{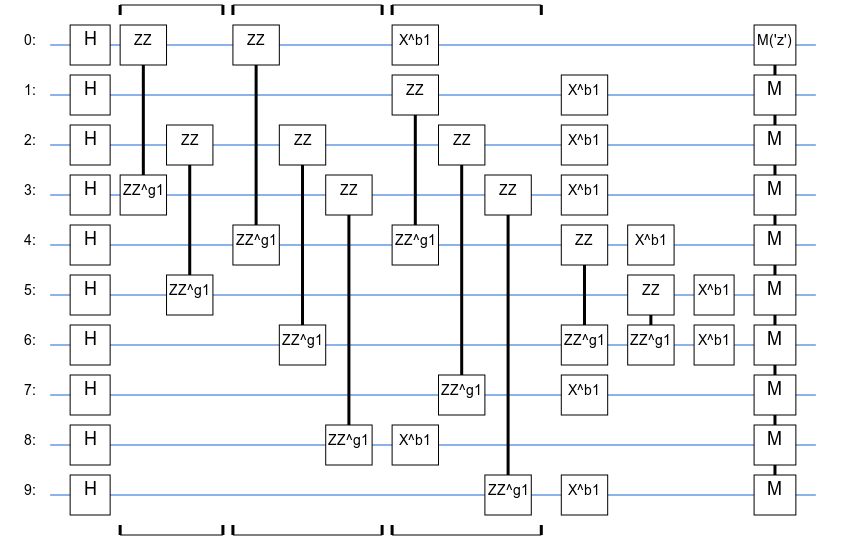} }}%
    \caption{QAOA for solving the MaxCut problem on random graphs.}
    \label{fig:qaoa-graph-circuit}
\end{figure}

\begin{figure}%
    \centering
    \subfloat[\centering Results across a sweep of circuit parametrizations.]{{\includegraphics[width=0.45\linewidth]{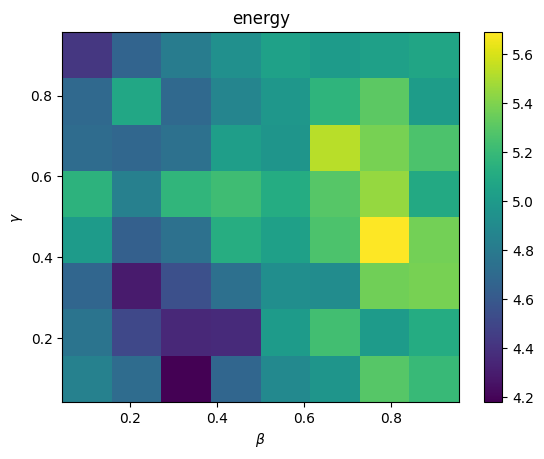} }}%
    \qquad
    \subfloat[\centering Target graph colored with final solution.]{{\includegraphics[width=0.45\linewidth]{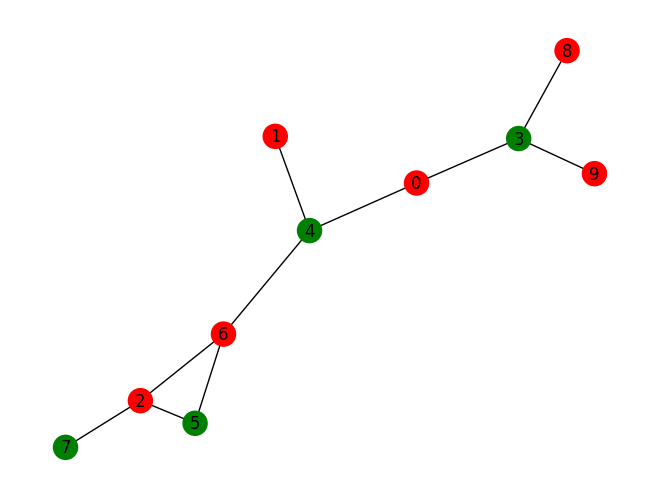} }}%
    \caption{The results of searching through possible circuit parametrizations, and the final MaxCut solution.}
    \label{fig:qaoa-results}
\end{figure}

Many realistic networks are large, random, and sparse, that is the degree of connectedness is much lower than the upper bound. These cases are particularly suited to matrix product states, as the corresponding QAOA circuit is very wide yet exhibits low entanglement and depth. More specific implementation details can be found in the \verb|bgls| documentation, and further information is available elsewhere in the literature \cite{Graham_2022}. We show here the results of using \verb|bgls| with MPS for optimizing such circuits and solving the MaxCut problem, as can be seen in figures \ref{fig:qaoa-graph-circuit} and \ref{fig:qaoa-results}.

First, a random Erdos-Renyi graph of 10 nodes and edge probability 0.3 is generated, and the corresponding QAOA circuit of 1 layer is constructed. For this problem, we use a custom subclass of \verb|cirq.contrib.quimb.MPSOptions| allowing restriction of the maximum degree of connectedness $\chi$ as described above.

The circuit is parameterized with $\gamma, \beta$, and an initial sweep of 100 samples is made for each configuration. The parameters maximizing average energy are chosen, and a final run of additional samples is done for this specific circuit. Of these, the bitstring sample that maximizes energy is chosen as the actual solution to the problem. For the graph seen in Fig.~7, the resultant energy (i.e. number of cuts) is 9, with the measured bitstring corresponding to the partition assignment of each node. The original graph is then colored with the solution and it can be confirmed to maximize the number of cuts.

This takes around 5 minutes on a laptop, with runtime coming primarily from tensor contractions in bitstring probability calculations. Hence, there is the possibility of much future improvement by investigating optimizations of the tensor network structure \cite{Gray2018}.

\section{Conclusion and Outlook}

We introduced \verb|bgls|, a Python package implementing the gate-by-gate algorithm for simulating sampling from quantum circuits~\cite{Bravyi_2022}. It interfaces with the Cirq framework to leverage its strengths and provide a familiar interface, yet remains highly flexible and extendable, functioning on any type of quantum state representation that can apply Cirq gates. 

We presented several useful representations \verb|bgls| supports out of box, namely stabilizer and matrix product states, covering their theoretical details as well as our specific implementation functionality. \verb|bgls| provides the relevant bitstring probability functions for these states as well as the more common state vector and density matrix representations, as well as the stochastic sum-over-Clifford technique for applying Clifford+$R_z(\theta)$ gates to stabilizer states. We then turned to the realistic example of solving MaxCut with QAOA using \verb|bgls|. 

BGLS is available to install directly via PyPI and its code is open source on GitHub. We hope BGLS is a useful tool for the community and welcome contributions.

\begin{acks}
AS thanks Giuseppe Carleo for hosting this project.
\end{acks}

\bibliographystyle{ACM-Reference-Format}
\bibliography{main-bib}

\end{document}